# INFLUENCE OF THE THEORETICAL MODEL EYE ON THE NUMERICAL EVALUATION OF FRACTAL INTRAOCULAR LENSES
# INFLUENCIA DEL MODELO DE OJO TEÓRICO EN LA EVALUACIÓN NUMÉRICA DE LENTES INTRAOCULARES FRACTALES

**Diego Montagud-Martínez[1,*], Vicente Ferrando[1], Juan A. Monsoriu[1] y Walter D. Furlan[2]**

*Corresponding Author: E-mail: diemonma@upvnet.upv.es

[1]*Centro de Tecnologías Físicas, Universitat Politècnica de València, 46022 Valencia, Spain*
[2]*Departamento de Óptica y Optometría y Ciencias de la Visión, Universitat de València, 46100 Burjassot, Spain*

## Abstract

In this work we present the numerical evaluation of a new design of fractal intraocular lens studied through a ray-tracing program. To determine the monochromatic and polychromatic performance of these lenses in different theoretical model eyes the Modulation Transfer Function (MTF) and the area above the MTF (AMTF) have been used. These merit functions show the same behavior for different values of asphericity (Q), independently from the theoretical model eye, even though there are differences due to the Spherical Aberration (SA) considered in each model.

## Resumen

En este trabajo presentamos la evaluación numérica de un nuevo diseño de lente intraocular fractal mediante un programa de trazado de rayos. Para determinar la respuesta tanto monocromática como policromática de estas lentes en diferentes modelos de ojos teóricos se ha utilizado la función de transferencia de modulación (Modulation Transfer Function, MTF) y el área bajo la MTF (AMTF). Estas funciones de mérito muestran el mismo comportamiento para diferentes valores de asfericidad (Q) independiente del modelo de ojo teórico, aunque existen pequeñas diferencias debido a la aberración esférica (Spherical Aberration, SA), considerada en cada modelo.

Descriptores: Lente intraocular; Fractal; Óptica.

## Introducción

Una lente intraocular (Intraocular Lens, IOL) es una lente que se inserta en el ojo sustituyendo al cristalino que se ha opacificado perdiendo su transparencia e impidiendo al ojo enfocar a diferentes distancias [1]. A este proceso quirúrgico, que es uno de los más empleados, se le denomina operación de cataratas. Además, debido al incremento de la esperanza de vida de las personas, este proceso de envejecimiento del cristalino va progresivamente en aumento [1]. Existen múltiples diseños de IOLs en el mercado. Dependiendo de la cantidad de focos que presenten tenemos lentes monofocales, con un foco para visión de lejos, bifocales, para visión de lejos y cerca, o trifocales con un foco adicional en distancia intermedia [2]. También existen lentes multifocales que además presentan una extensión de la profundidad de foco utilizando como mecanismo la SA. Este tipo de lentes se les conoce como lentes de foco extendido (Extended Depth of Focus, EDOF) [3].

A la hora de diseñar una nueva IOL es imprescindible en primer lugar realizar simulaciones numéricas con el modelo de la nueva lente (con programas de trazado de rayos). Posteriormente y una vez fabricada la IOL se mide experimentalmente las propiedades ópticas de la misma en el banco óptico y finalmente se realizan ensayos clínicos con pacientes. Para las simulaciones numéricas se utilizan programas de trazado de rayos que permiten evaluar la calidad óptica de cualquier elemento óptico, en este caso de una IOL. Dependiendo de la complejidad de la simulación se pueden obtener resultados más reales utilizando modelos de ojo teóricos propuestos previamente en la bibliografía [4]. Existen múltiples modelos de ojos teóricos y cada uno de ellos tiene sus ventajas y desventajas. Entre los más destacados encontramos los modelos de Atchison [5] y Liou-Brennan [6], cuyos parámetros oculares están basados en datos biométricos de pacientes con una edad media de 45 años, o el de Navarro [7]. Estos tres modelos presentan dos superficies corneales, una pupila y cristalinos de diferente índole. Sin embargo, los parámetros oculares para los distintos modelos discrepan y las mayores diferencias se encuentran en las Q corneales que afectan directamente a la SA del modelo de ojo, así como el ángulo de incidencia de la luz y el índice de refracción del cristalino.

El objetivo de este trabajo es evaluar numéricamente mediante el programa de trazado de rayos ZEMAX™ OpticStudio (EE versión 18.7, ZEMAX Development Corporation, Bellevue, Washington, USA) tres modelos de ojos teóricos en los que se ha extraido el cristalino y se ha introducido una nueva IOL bifocal EDOF de diseño propio basada en estructuras fractales [8].

## Métodos

### Nuevo diseño de lente intraocular fractal

La lente estudiada en este trabajo consiste en un diseño híbrido refractivo-difractivo [8] el cual proporciona una extensión de foco y una baja aberración cromática. Está inspirada en la placa zonal fractal de Cantor [9]. En la Figura 1 se puede observar como a partir del conjunto triádico de Cantor de orden 2 definido a lo largo de la coordenada radial cuadrática se obtiene el perfil fractal de la lente propuesta. La IOL resultante presenta una distribución fractal de zonas anulares de diferente potencia (radio de curvatura) generando un diseño refractivo-difractivo con dos focos principales uno, para visión de lejos y otro para visión de cerca. En el software de trazado de rayos se introdujo la nueva lente intraocular fractal (Fractal Intraocular Lens, FIOL) como una lente de dos superficies ($r_{ant}$= 21.4 mm y $r_{post}$= -17.0 mm) con un espesor de 0.7 mm y un índice de refracción (n=1.55) similar al de las IOL comerciales. El perfil fractal se introdujo en ZEMAX en la cara anterior como una superficie "grid sag", en la cual se introducen las sagitas de la superficie fractal, mientras que la cara posterior de la FIOL presentaba un diseño esférico (Q=0) o asférico (Q=-10).

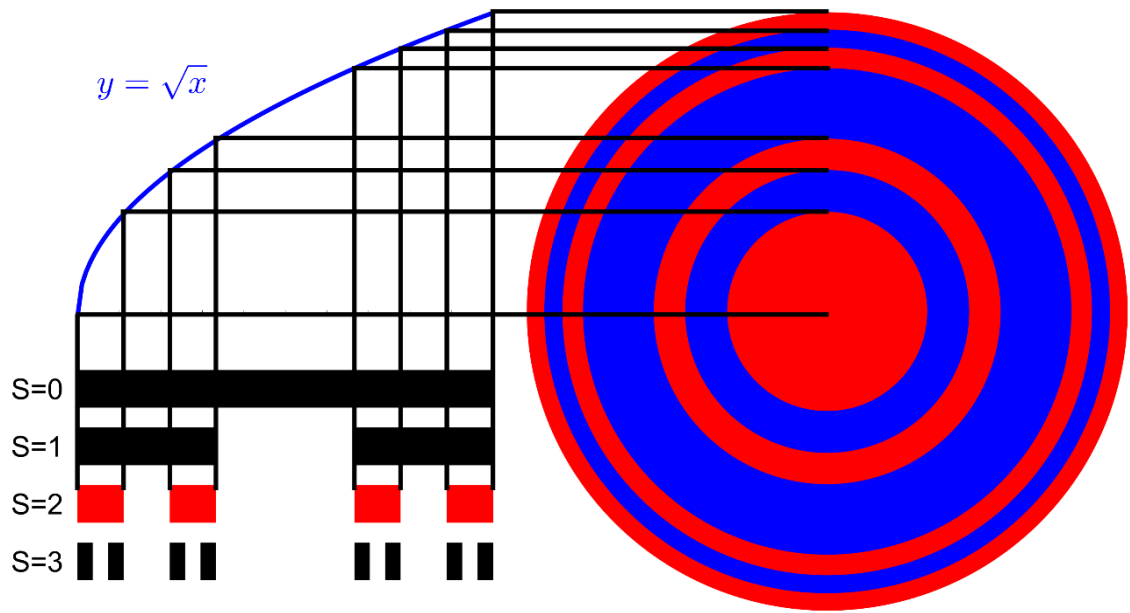

*Figura 1: Obtención del perfil fractal de la lente a partir del conjunto triádico de Cantor.*

**Modelos de ojos teóricos**

Para este estudio se utilizaron tres ojos teóricos ampliamente utilizados: Atchison [5], Liou-Brennan [6] y Navarro [7]. En la tabla 1 se muestran los parámetros oculares de cada modelo. No se muestran los datos del cristalino ya que en este estudio se sustituyeron por la FIOL. El valor de Q para la cara anterior y posterior de la córnea en cada modelo de ojo afecta al valor global de la SA de dicho modelo. Esto unido a la FIOL esférica (Q=0) o asférica (Q=-10) generará diferencias entre los modelos de ojo.

|  |  | **Atchison** | **Liou-Brennan** | **Navarro** |
|---|---|---|---|---|
| Córnea Anterior | Radio curvatura (mm) | 7.77 | 7.77 | 7.72 |
|  | Q | -0.15 | -0.18 | -0.26 |
|  | n | 1.376 | 1.376 | 1.3771 |
|  | Espesor (mm) | 0.55 | 0.55 | 0.55 |
| Córnea Posterior | Radio curvatura (mm) | 6.40 | 6.40 | 6.5 |
|  | Q | -0.275 | -0.60 | 0 |
|  | n | 1.3374 | 1.336 | 1.3374 |
|  | Espesor (mm) |  | 3.16 | 3.05 |
| FIOL anterior (perfil fractal) | Radio curvatura (mm) | 21.40 | | |
|  | Q | 0 | | |
|  | n | 1.55 | | |
|  | Espesor (mm) | 0.70 | | |
| FIOL posterior | Radio curvatura (mm) | -17.00 | | |
|  | Q | 0 o -10 | | |
|  | n | 1.336 | 1.336 | 1.336 |
|  | Espesor (mm) | Ajustado | | |

*Tabla 1: Parámetros oculares de los tres modelos utilizados.*

**Registro de medidas**

Después de ajustar la longitud axial de cada modelo de ojo se procedieron a medir las MTFs de los tres modelos con la IOL sin el perfil fractal (monofocal), con la FIOL esférica y con la FIOL asférica en las vergencias comprendidas entre +0.50 D y -3.50 D en pasos de 0.25 D. La luz utilizada para la obtención de las MTFs fue monocromática ($\lambda$=555 nm) y policromática siguiendo la distribución V($\lambda$) dada por el propio software ($\lambda_1$=470 nm, $\lambda_2$=510 nm, $\lambda_3$=555 nm, $\lambda_4$=610 nm, $\lambda_5$=650 nm con sus respectivos pesos de 0.091, 0.503, 1, 0.503 y 0.107). Las pupilas que se utilizaron en las simulaciones fueron de 3.0 mm y 4.5 mm. El iris se situó en el mismo plano de la cara anterior de la IOL para los tres modelos. Utilizando un código propio de Matlab (versión R2018b, Mathworks Inc, Natick, Massachusetts, USA) se calcularon por el método de los trapecios el área bajo la curva de cada MTF (AMTF) entre las frecuencias de 9.49 ciclos/grado y 59.86 ciclos/grado para cada uno de los modelos en las condiciones antes descritas. Dichas frecuencias equivalen a agudezas visuales de 0.32 y 2.00 en escala decimal. Nótese que esta métrica, AMTF, se correlaciona muy bien con la AV correspondiente [10]

## Resultados y discusión

En la Figura 2 se muestran las MTFs de los focos de lejos para pupilas de 3.0 mm (a) y 4.5 mm (b), respectivamente. Como puede observarse para los tres modelos de ojos no existen muchas diferencias entre la luz monocromática y la luz policromática. Por tanto, la aberración cromática de los modelos para ambas condiciones lumínicas es despreciable. También se observa que en el modelo de Atchison y Navarro las MTFs son más altas, tanto la monofocal como las FIOLs. En el caso del modelo de Liou-Brennan las MTFs son menores pero la diferencia entre la monofocal y las FIOLs en proporción es menor que para los otros dos modelos. El comportamiento de las lentes es más similar en los modelos de Atchison y Navarro que en el modelo de Liou-Brennan debido a la discrepancia y complejidad de este último con respecto a los dos anteriores. Además, la diferencia entre la FIOL esférica y la FIOL asférica aumenta al aumentar la pupila debido a que la influencia de la Q y la SA global del ojo es mayor cuanto mayores son las distancias de los rayos no paraxiales con respecto al eje visual.

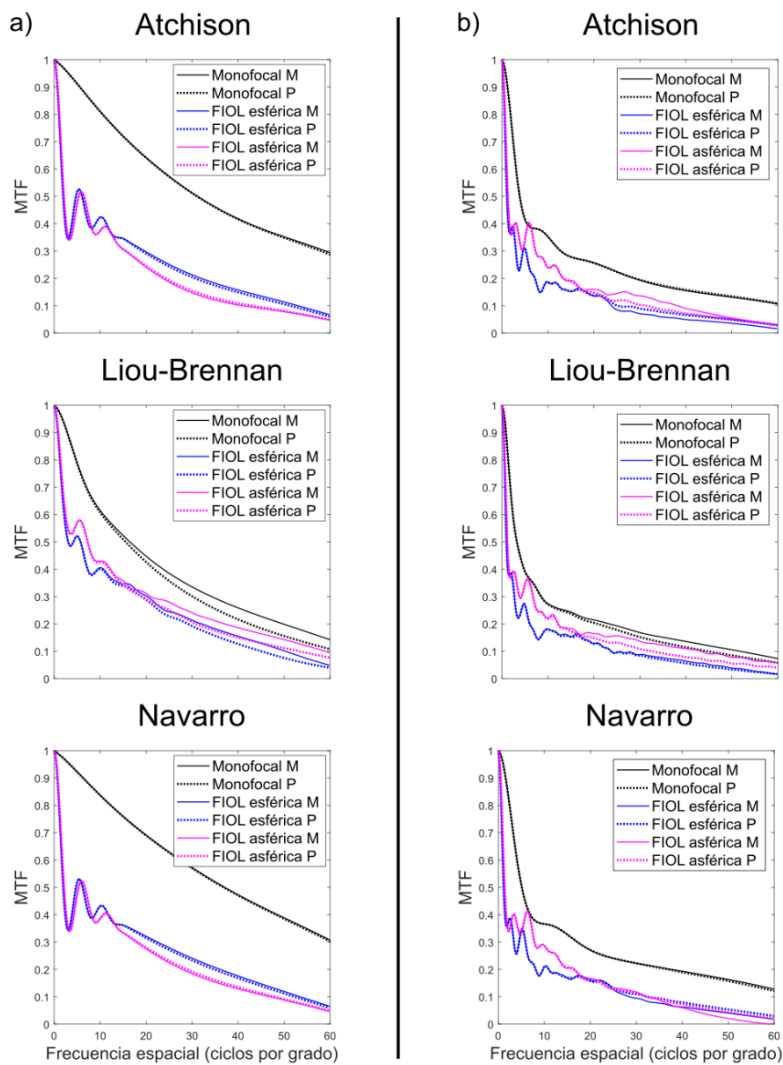

*Figura 2: MTFs en el foco de lejos para pupila de 3.0 mm (a), MTFs en el foco de lejos para pupila de 4.5 mm (b). Las líneas continuas representan luz monocromática y las discontinuas luz policromática.*

En la Figura 3 se observan las MTFs de los focos de cerca para pupilas de 3.0 mm (a) y 4.5 mm (b). Nótese como las MTFs en estas figuras están en escala logarítmica para poder observar mejor las gráficas. Al igual que ocurre con el foco de lejos, los modelos de Atchison y Navarro predicen valores de MTFs más similares que en el caso del de Liou-Brennan. Además, también se confirma que al aumentar la pupila las diferencias entre la FIOL esférica y asférica aumentan. Al incrementar el tamaño pupilar la luz más alejada del eje visual no atraviesa el perfil fractal y por tanto focaliza en lejos. Por eso, las MTFs en 3.0 mm son mejores que en 4.5 mm sobre todo para las altas frecuencias.

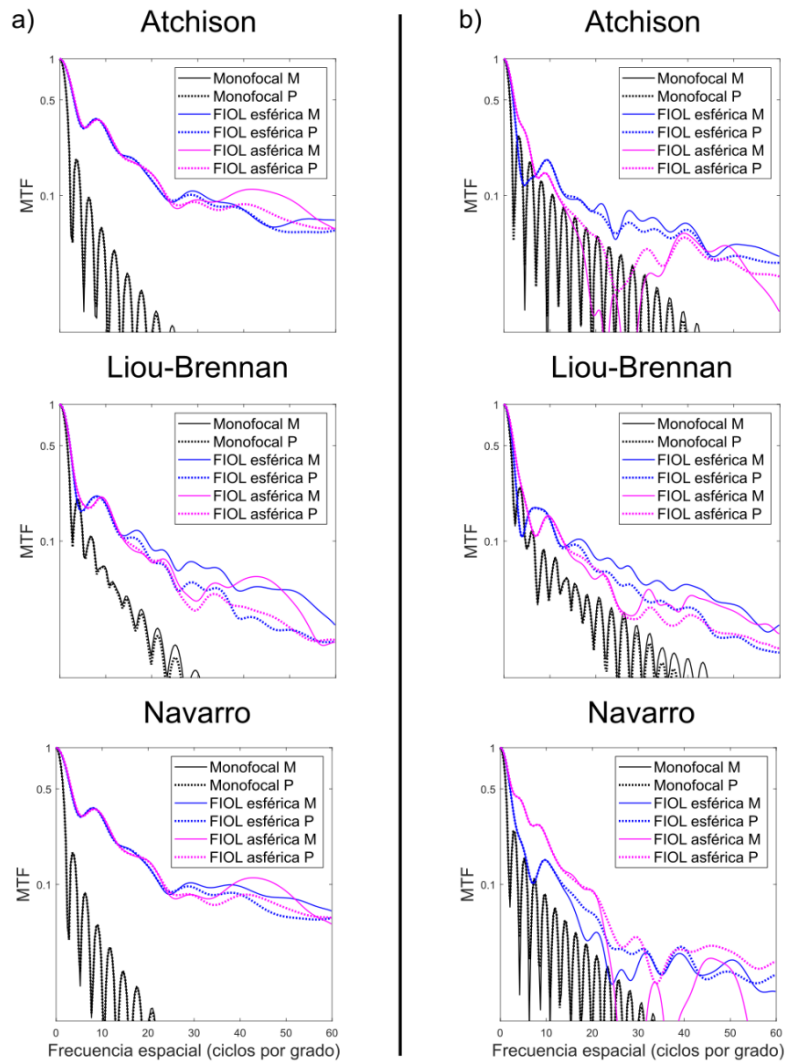

*Figura 3: MTFs en el foco de cerca para pupila de 3.0 mm (a), MTFs en el foco de cerca para pupila de 4.5 mm (b). Las líneas continuas representan luz monocromática y las discontinuas luz policromática.*

A pesar de que las MTFs muestran buenos resultados, las AMTFs a través de foco permiten observar el comportamiento de la lente en diferentes vergencias tal y como se muestra en la Figura 4 para pupilas de 3.0 mm (a) y 4.5 mm (b), respectivamente. En la Figura 4a se muestra claramente el perfil bifocal de la FIOL y como no existen diferencias significativas entre las medidas con luz monocromática y luz policromática. Esto último puede comprobarse también en la Figura 4b. Centrándonos en la pupila de 3.0 mm observamos como para los tres modelos de ojo el efecto de añadir una superficie asférica al perfil fractal es el mismo, es decir, desplaza los focos hacia vergencias positivas, si bien este desplazamiento es ligeramente mayor para el foco de cerca. Además, en los tres casos las alturas de las AMTFs en lejos son mejores para la FIOL asférica que la esférica.

Si nos fijamos en la Figura 4b podemos observar como las AMTFs caen de manera significativa debido a las aberraciones de los modelos. Es decir, al aumentar el tamaño pupilar a 4.5 mm los rayos que llegan a la retina dejan de ser paraxiales provocando peores AMTFs tanto para la monofocal como para ambas FIOLs. Además, se produce una la extensión del foco de cerca como consecuencia de la estructura fractal del diseño. También se puede observar como la inclusión de una superficie asférica a la FIOL mejora el foco de lejos y de cerca en los tres modelos de ojo teórico.

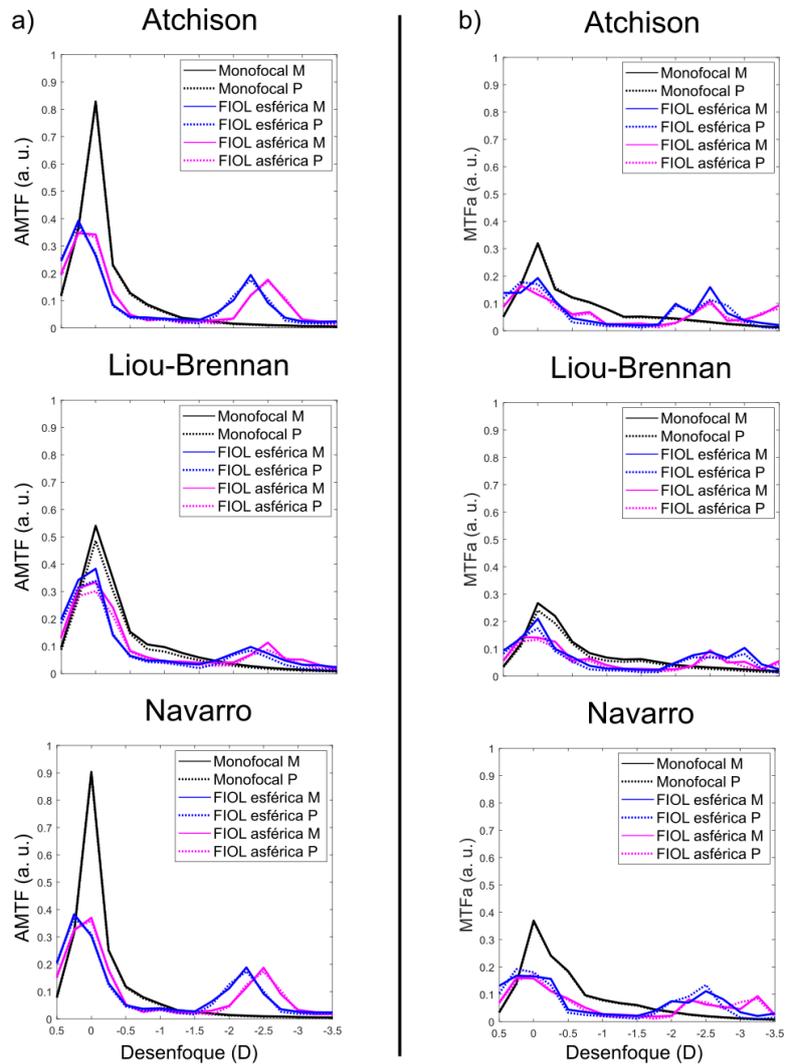

*Figura 4: AMTFs para pupila de 3.0 mm (a), AMTFs para pupila de 4.5 mm (b). Las líneas continuas representan luz monocromática y las discontinuas luz policromática.*

## Conclusiones

Los programas de trazado de rayos constituyen una herramienta esencial para la validación numérica de nuevos diseños de elementos ópticos. Estos programas permiten determinar la calidad óptica de estos elementos mediante funciones de mérito como son la MTF y la AMTF, e incluso simular imágenes. En este trabajo se ha caracterizado una lente con perfil fractal y con dos asfericidades diferentes para tres modelos de ojos teóricos. Los resultados muestran que, aunque los modelos de ojos teórico presentan resultados diferentes, la tendencia es similar para los tres. La FIOL propuesta presenta un perfil bifocal con foco extendido en cerca para todos los modelos de ojo que mejora al añadir una superficie asférica.

**Bibliografía**

[1] W. N. Charman, Ophthalmic Physiol. Opt. **34** (2014) 297.

[2] K. Hayashi, S. I. Manabe & H. Hayashi, J Cataract Refract Surg **35** (2009) 2070.

[3] J. Liu, Y. Dong, & Y. Wang, BMC Ophthalmol **19** (2019) https://doi.org/10.1186/s12886-019-1204-0.

[4] D.A. Atchison, Schematic eyes, 1st Ed. (Taylor & Francis, Boca Raton, 2017), pp 235-247.

[5] David A. Atchison, Vision Res. **46** (2006) 2236.

[6] H. Liou & N. A. Brennan, J. Opt. Soc. Am. A. **14** (1997) 1684.

[7] I. Escudero-Sanz & R. Navarro, J. Opt. Soc. Am. A. **16** (1999) 1881.

[8] L. Remón, S. García-Delpech, P. Udaondo, V. Ferrando, J. A. Monsoriu & W. D. Furlan, PLoS ONE **13** (2018) https://doi.org/10.1371/journal.pone.0200197.

[9] G. Saavedra, W. D. Furlan & J. A. Monsoriu, Opt. Lett. **12** (2003) 971.

[10] A. Alarcon, C. Canovas, R. Rosen, H. Weeber, L. Tsai, K. Hileman, & P. Piers, P. Biomedical Optics Express, **7** (2016) 1877.